\documentstyle[seceq,epsf,wrapft]{ptptex}

\newcommand{\be}{\begin{equation}}
\newcommand{\ee}{\end{equation}}
\newcommand{\ba}{\begin{array}{c}}
\newcommand{\ea}{\end{array}}
\newcommand{\bqa}{\begin{eqnarray}}
\newcommand{\eqa}{\end{eqnarray}}
\notypesetlogo  
\title{How to parameterize a light and broad resonance (the $\sigma$
meson)\footnote{Talk given at symposium on $Hadron$
$Spectroscopy$, $Chiral$ $Symmetry$ $and$ $Relativistic$
$Description$ $of$ $Bound$ $Systems$, Nihon University, Tokyo,
Japan. Feb. 24 -- 26, 2003 } }
\author{%
Hanqing {\sc Zheng} } \inst{Department of Physics, Peking
University, Beijing 100871, P.~R.~China }
\recdate{%
\today  }
\abst{%
  We point out that  a commonly used
parameterization form to describe the $\sigma$ resonance is
problematic by introducing a spurious singularity below two
particle threshold, on the second sheet. The spurious singularity
violates chiral symmetry and leads to sizable contribution when
the resonance is light and broad -- as what happens to the
$\sigma$ resonance,  hence  must be removed.}

\begin{document}
\maketitle

\setcounter{tocdepth}{4}

The evidence for the  existence of a light and broad resonance in
I=J=0 $\pi\pi$ scattering process -- named as the $\sigma$
resonance -- has been accumulated, and more and more physicists
working on the subject accepted the renaissance of the $\sigma$
resonance. Among those works, model independent
analysis~\cite{XZ00,CGL01} are vital in clarifying the issue
whether the $\sigma$ resonance exists or not, since one has to
distinguish and separate the contribution of a broad resonance to
the phase motion from background contributions, and the latter is
often very difficult, if not impossible, to estimate. However,
there are different opinions in the literature against the
existence of the light $\sigma $ resonance~\cite{MO02}, based upon
the following observation: the $\pi\pi$ phase shift does not pass
$90^\circ$ at low energies, which would have occurred, according
to Eq.~(\ref{BW2}) given below, at $s=M^2$ with $M$ the
Breit--Wigner mass which should be light. This question is
actually already answered, in one aspect, by Ishida and
collaborators.~\cite{Ishida}  who pointed out that chiral symmetry
requires the background contribution to the phase shift in I=J=0
$\pi\pi$ scatterings to be negative,\footnote{The negative
background contribution  is found  in chiral perturbation theory
($\chi PT$) in Ref.~\cite{XZ00}.} Ishida and collaborators argue
that the background phase cancels the large phase motion generated
by the Breit--Wigner resonance and therefore there is no
contradiction between the observed $\pi\pi$ phase shift and the
existence of a light and broad resonance. In this talk I will
further point out that a conventional Breit-Wigner description to
the $\sigma$ resonance actually violates chiral symmetry and
should be abandoned, and it is not even necessary for a light and
broad resonance to develop a phase shift passing $90^\circ$.

\section{ The Breit--Wigner formula as a narrow width approximation}
\vspace{-0.5cm}

In principal the Breit--Wigner description of resonance only works
for infinitely small width. This is clearly understood from the
following heuristic derivation of the Breit--Wigner
formula.~\cite{gao}  First of all, for the propagator of a stable
particle we have
 { \begin{equation}\label{stab}
\triangle(p^2)=\frac{1}{p^2-m^2+i\epsilon}\ . \end{equation}}
 Now let us focus on the following chain decay process, {\begin{equation}\label{ABC}
 A \to B + C, \,\,\,   B \to D + E\ ,
 \end{equation}¡¡
 described by the effective Hamiltonian $H_{eff} = f A B C + g B
 D E$ where $D$ and $E$ are massles particles and $B$ is a narrow resonance. The  decay
 of particle $A$ can be viewed either  as a two body decay process, $A\to
 B+C$, which has a decay width\footnote{Assuming $B$ is stable in obtaining $\Gamma_A$.
 The decay width of $B$ is also
 obtainable: $\Gamma_B=\frac{g^2}{16\pi}\frac{1}{m_B}$}
 { \be
 \Gamma_A=\frac{f^2}{16\pi}\frac{m_A^2-m_B^2}{m_A^3}\ ,\,\,\, 
     \ee}}
     or as a three body decay, $A\to B+C+E$. To calculate the 3
     body decay we need a knowledge on the form of the  propagator
     of the unstable particle $B$. By comparing with
     Eq.~(\ref{stab}) we may assume the propagator to be the following
 form:
 {\be\label{unstab} \triangle_B(p^2)=\frac{1}{p^2-\alpha+i\beta}\ ,
 \ee}
where $\alpha$ and $\beta$ are constants. Now the partial decay
width of the process $A\to C+D+E$ can  be calculated since we know
the effective Hamiltonian and Eq.~(\ref{unstab}) (in the
calculation $m_A
>> \Gamma_A$ is assumed!),
{\be\label{BWgao}
\Gamma_A=\frac{f^2}{16\pi}\frac{g^2}{16\pi}\frac{m_A^2-\alpha}{m_A^3\beta}
.\ee } The two different calculation of $\Gamma_A$ have to yield
the same result, which leads to
 {  \be\label{SBW}
\triangle_{BW}(p^2)=\frac{1}{p^2-m_B^2+im_B\Gamma_B}\ . \ee}
 This equation is the well known Breit--Wigner formula.
From the way we derive it we realize that the Breit--Wigner
description of resonances can only be exact in the case of
infinitely small width, and one expects that it works reasonably
well in narrow width approximation. However, there is a problem
associated with Eq.~(\ref{SBW}), since it missed the threshold
effect and the parameterization form does not respect a nice
property of Feynman amplitudes called real
analyticity.\footnote{The latter is not crucial but the former is
really a shortcoming of the standard Breit--Wigner
parameterization.} Therefore, one often adopts another
parameterization form instead of Eq.~(\ref{SBW}),
  { \be\label{BW2}
\triangle_{BW}'(p^2)=\frac{1}{p^2-M^2+i\rho(s)G}\ , \ee} where
$\rho(s)$ is the kinematic factor. For equal mass case like
$\pi\pi$ scattering
 it is $\rho(s)=\sqrt{1-4m_\pi^2/s}$. Both Eq.~(\ref{SBW}) and
 Eq.~(\ref{BW2}) are frequently used in the literature and are recommended by the
 Particle Data Group. It is worth emphasizing that when the
 resonance's width is narrow and the pole locates far above the
 threshold ($M^2>>4m_\pi^2$), in the vicinity of $s=M^2$ the
 Eq.~(\ref{SBW}) and Eq.~(\ref{BW2}) gives almost identical
 results. It is noticed that in more complicated applications people sometimes also
 treat the constant $G$ in Eq.~(\ref{BW2}) as a function of $s$.

 The condition for applying the Breit--Wigner approximation is actually well
known among  physicists working in the field, what I will focus on
in this talk is only the $\sigma $ resonance, which is light and
broad~\cite{XZ00,CGL01,XZ03}. Therefore one hesitates to use the
parameterization form like  Eq.~(\ref{BW2}) in the experimental
fit to extract the $\sigma$ pole position. Unfortunately this is
not the case in many of the research works.  Before demonstrating
Eq.~(\ref{BW2}) should $not$ be used to parameterize the $\sigma$
resonance, it is necessary to study some general properties of the
scattering $S$ matrix.
\section{The factorized $S$ matrix}
\subsection{The simplest partial wave $S$ matrices} \label{TfSm}
The starting
point is the dispersion relations set up by the present author and
collaborators on  $\pi\pi$ scattering partial wave $S$ matrix
elements~\cite{XZ00,XZ01},
 \bqa
 &&\sin(2\delta_\pi)\equiv\rho F\ ,\nonumber \\
  F(s)&=& \alpha-\sum_j{1/ 2i\rho(z^{II}_j)\over
 S'(z^{II}_j)(s-z^{II}_j)}  +{1\over\pi}\int_L{{\rm Im}_LF(s')
  \over s'-s} ds'+{1\over\pi}\int_R{{\rm Im}_RF(s')
  \over s'-s} ds',
   \label{sin2d}\eqa
  in which we did not include the bound state contribution for simplicity, since it is irrelevant here.
   In the above expression
    $L=(-\infty,0]\ , R=[4m_K^2,\infty)$, $\alpha$ is a subtraction constant and
   $z_j$ denotes the pole position on the second sheet. When $z_j$ is real it represents a virtual state
   pole, when $z_j$ is complex it must appear in one pair together with $z_j^*$, representing a resonance.
   The experimental curve of the function $F$ is convex, yet chiral perturbation theory predicts
   a negative and concave left hand integral contribution.\cite{XZ00} This fact undoubtfully establish
   the existence of the $\sigma$ resonance. Similarly we have~\cite{HXZ02},
 {\bqa
   \cos(2\delta_\pi) \equiv\tilde{F}
   &&=\tilde{\alpha}+\sum_j{1\over
 2S'(z^{II}_j)(s-z^{II}_j)}
+{1\over\pi}\int_L {{\rm
 Im}_L\tilde F(s')
 \over s'-s} ds' \nonumber\\ &&+{1\over\pi}\int_R {{\rm
 Im}_R\tilde F(s')
 \over s'-s} ds'\ . \label{cos2d}
\eqa}
 Eqs.~(\ref{sin2d}) and (\ref{cos2d}) defines the analytic
continuation of the scattering $S$ matrix,
$S=\cos(2\delta_\pi)+i\sin(2\delta_\pi)$ and further, the
generalized unitarity equation,
 {\be
\sin^2 2\delta_\pi+\cos^2 2\delta_\pi \equiv 1\ ,
 \label{constraint}
 \ee} which is valid
on the whole complex $s$ plane.

In general the $S$ matrix is very complicated since it may contain
various poles and cuts. The latter, according to the discussion
given above can either be originated from the kinematic factor, or
can be generated by those dispersion integrals in
Eqs.~(\ref{cos2d}) and (\ref{sin2d}). Here I call all those cuts
generated from the left hand integrals  'dynamical' cuts, though
such a terminology may be somewhat misleading.

Before studying more complicated case of $S$ matrix in reality,
let us first focus on the those simplest solutions of the $S$
matrix. Here 'simplest' means there is no 'dynamical' cut and the
number of poles is minimal. In the absence of bound state there
are two such  simplest solutions (taking the mass of the
scattering particle to be unity):
\begin{enumerate}
\item A virtual state pole at $s=s_0$ ($0<s_0<4$: scattering length {$a=\sqrt{s_0\over 4-s_0}$}.
 {\bqa\label{SR1}
 {\rm Re}_RT(s)&=&\frac{1}{4}\sqrt{s_0(4-s_0)}\frac{s}{s-s_0}\ ,\nonumber\\
{\rm Im}_R T(s)&=&\rho(s)\frac{ss_0}{4(s-s_0)}\ .
 \eqa}
or the $S$ matrix ,{ \be S^v(s)=\frac{1+i\rho(s)a }{1-i\rho(s)a}\
. \ee}
\item A (pair of) resonance:
poles locate at $z_0$ and $z_0^*$ on the second sheet. The
solution is:
 \bqa\label{SR1''}
 {\rm Re}_RT(s)&=&\Delta {\rm
 Re}[\sqrt{z_0(z_0-4)}]\frac{s(r_0-s)}{(s-z_0)(s-z_0^*)}\ ,\nonumber\\
{\rm Im}_R T(s)&=&\Delta {\rm
Im}[z_0]\rho(s)\frac{s^2}{(s-z_0)(s-z_0^*)}\ ,
 \eqa
 where \bqa\label{r_0}
 \Delta=\frac{{\rm Im}[z_0]}{({\rm
Re}[\sqrt{z_0(z_0-4)}])^2+({\rm Im}[z_0])^2}\ ,\,\,\,
r_0[z_0]={\rm Re}[z_0] + {\rm Im}[z_0]\frac{{\rm
Im}[\sqrt{z_0(z_0-4)}]}{{\rm Re}[\sqrt{z_0(z_0-4)}]}\ .\nonumber
\eqa
 The $S$ matrix can be rewritten as,
  {\be\label{sres}
S^R(s)=\frac{r_0[z_0]-s+i\rho(s)s\frac{{\rm Im}[z_0]}{{\rm
Re}[\sqrt{z_0(z_0-4)}]}}{r_0[z_0]-s-i\rho(s)s\frac{{\rm
Im}[z_0]}{{\rm Re}[\sqrt{z_0(z_0-4)}]}}\ , \ee } where { \be
r_0[z_0]={\rm Re}[z_0] + {\rm Im}[z_0]\frac{{\rm
Im}[\sqrt{z_0(z_0-4)}]}{{\rm Re}[\sqrt{z_0(z_0-4)}]}\ .\ee}
 This solution is unique
for the $S$ matrix with only one resonance  and without the so
called `dynamical cuts'.
\end{enumerate}

\begin{figure}[htb]
\parbox{\halftext}
{ \epsfysize=4cm \centerline{\epsffile{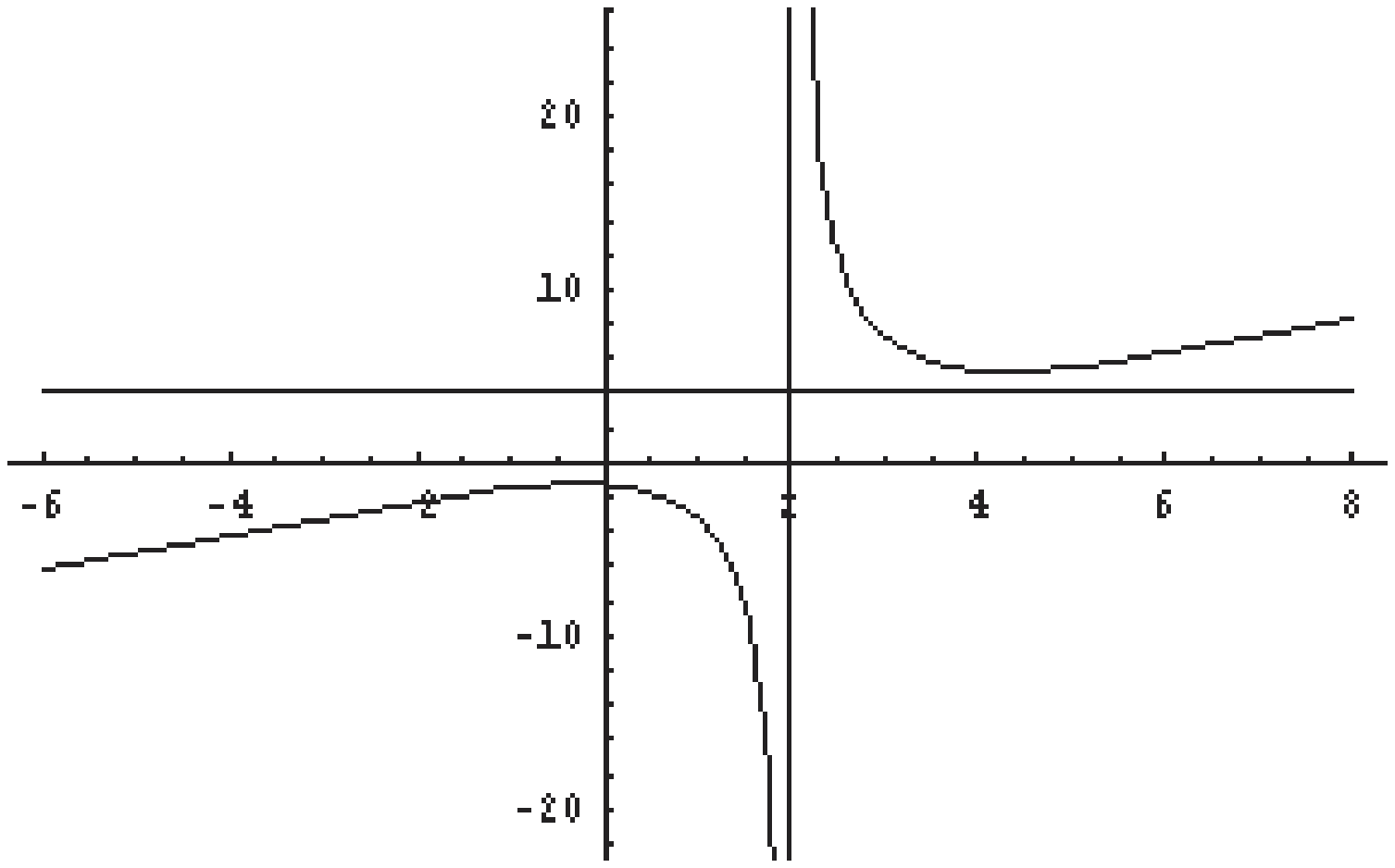}}
\caption{$r_0[z_0]$ as a function of ${\rm Re}[z_0]$, fixing
 ${\rm Im}[z_0]=1$. The vertical line corresponds to ${\rm Re}[z_0]=2$ and the horizontal line
 corresponds to $r_0=4$.}
\label{figs0}}
 \hspace{8mm}
\parbox{\halftext}
{ \epsfysize=4cm \centerline{\epsffile{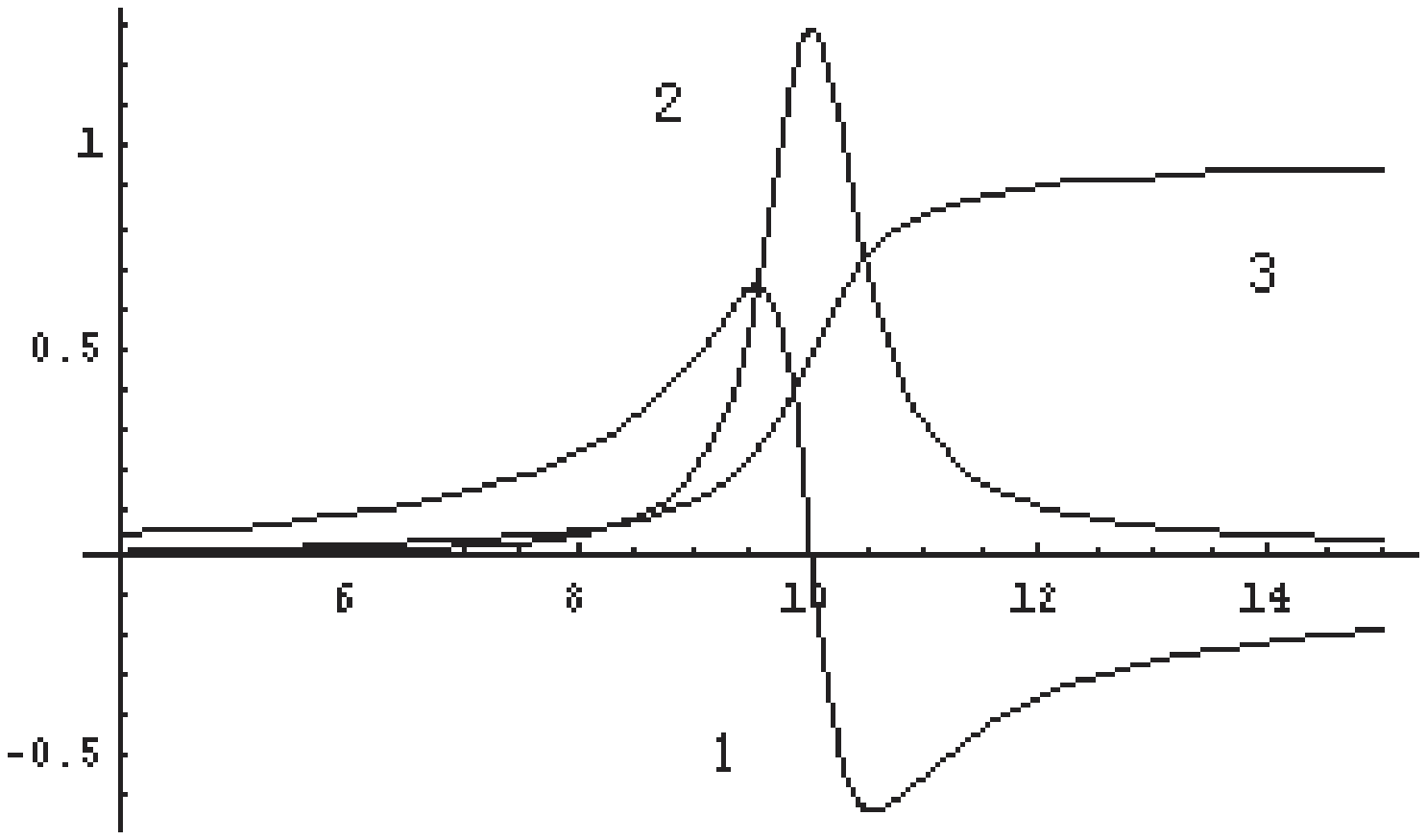}} \caption{The
qualitative behavior of a narrow resonance with ${\rm
Re}[z_0]=10$, ${\rm Im}[z_0]=.5$. Line 1 represents the real part
of the $T$ matrix, line 2 represents the imaginary part of the $T$
matrix whereas  line 3 represents the phase shift.
}\label{figre1}}
\end{figure}

\begin{figure}[htb]
\parbox{\halftext}
{ \epsfysize=4.5cm \centerline{\epsffile{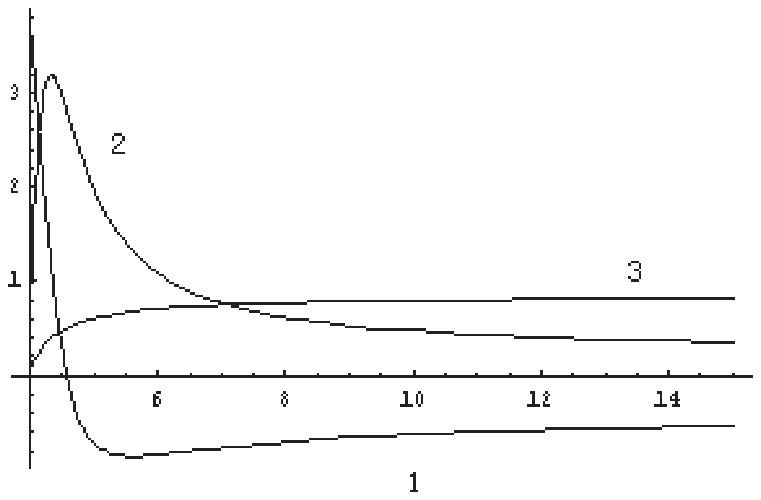}} \caption{The
behavior a resonance with  ${\rm Re}[z_0]=4$, ${\rm Im}[z_0]=.5$.
Notations 1,2 and 3 in the figure have the same meaning as in
fig.~\ref{figre1}. } \label{figre2}}
 \hspace{8mm}
\parbox{\halftext}
{ \epsfysize=4.5cm \centerline{\epsffile{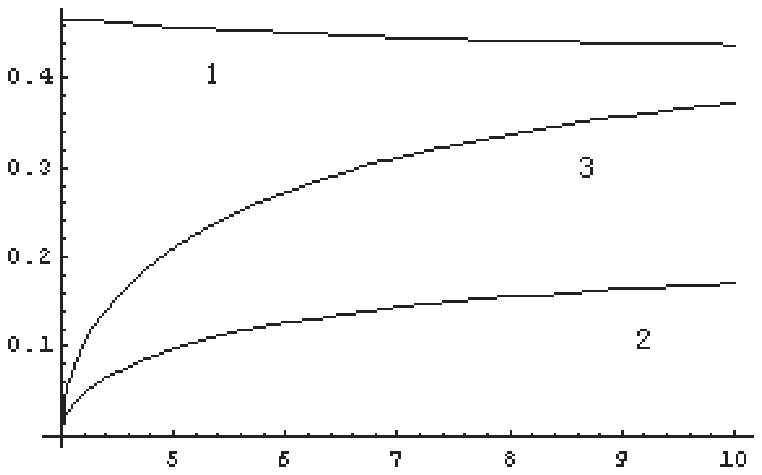}} \caption{The
behavior of a resonance with  ${\rm Re}[z_0]=0$, ${\rm
Im}[z_0]=.5$. Notations 1,2 and 3 in the figure have the same
meaning as in fig.~\ref{figre1}. }\label{figre3}}
\end{figure}

It is very interesting to see $r_0$ as a function of $z_0$, as
shown in Fig.~\ref{figs0}. When ${\rm Re}[z_0]>>{\rm Im}[z_0]$,
$r_0$ is very close to ${\rm Re}[z_0]$ and the result of
Eq.~(\ref{sres}) is similar to the Breit--Wigner formula
Eq.~(\ref{BW2}), as can be seen from Fig.~\ref{figre1}. However
when decreasing ${\rm Re}[z_0]$ while keeping ${\rm Im}[z_0]$
fixed $r_0[z_0]$  does not decrease monotonously.
On the contrary, it increases when ${\rm Re}[z_0]$ gets small
enough. When the resonance becomes light and broad, the typical
phase motion it exhibits is depicted in Fig.~\ref{figre2}. That is
the phase increases slowly and reach $90^\circ$ at very distant
place. 
I believe, this simple picture as shown by Fig.~\ref{figre2},
nicely reveals qualitatively what happens in I=J=0 $\pi\pi$
scattering at low energies\footnote{and even $\pi K$ scattering at
low energies, though  in $\pi K$ scattering the singularity
structure of the \protect{$S$} matrix is more complicated due to
unequal mass kinematics.}. Since $s=r_0$ is the point where the
phase pass $90^\circ$, Eq.~(\ref{sres}) characterizes the drastic
difference between the phase motion generated by a light and broad
resonance and a narrow resonance located far away from the
threshold.

There is a critical  line corresponds to ${\rm Re}[z_0]=2$ on the
$s$ plane. When the resonance locates on the right hand side of
the line, i.e.,  ${\rm Re}[z_0]>2$, the phase shift can get larger
than $\pi/2$; whereas when the resonance locates on the left hand
side of the line, i.e.,  ${\rm Re}[z_0]<2$, the phase shift can
never reach $\pi/2$. Of course, a resonance will always give a
positive contribution to the phase shift increasing monotonously
as (the physical value) $s$ increase. But a deeply bounded
resonance corresponding to ${\rm Re}[z_0]<<2$ behaves  like a
normal virtual state.
\vspace{1cm}
\subsection{The violation of Levinson's theorem }

In the examples discussed above, the phase shift of a virtual
state is \be\label{rphase} \tan
(\delta_v)=\rho(s)\sqrt{\frac{s_0}{4-s_0}}\ . \ee
 For the phase
shift of a resonance, it is \be\label{rphase''} \tan
(\delta_r)=\frac{{\rm Im}[z_0]}{{\rm
Re}[\sqrt{z_0(z_0-4)}]}\frac{\rho(s)s}{r_0-s}\ . \ee
 If $r_0>4$
then the phase shift passes $90^\circ$ when $s$ passes $r_0$. At
$s=\infty$, \be\label{del1}
\delta(\infty)=\pi-\tan^{-1}(\frac{{\rm Im}[z_0]}{{\rm Re}[\sqrt
{z_0(z_0-4)}]} )< \pi \ .\ee This disagrees with Levinson's
theorem in scattering theory, where the theorem says that a
resonance contributes a phase shift $180^\circ$ at $\infty$, The
discrepancy comes from the fact that in here we have the left hand
cut from the relativistic kinematics, which leads to physical
effects violating  Levinson's theorem.
 Actually, here we have
{ \be\label{del1'} \delta(\infty)-\delta(-\infty)=\pi\ .\ee}
\subsection{The factorized $S$ matrix}
In above we have analyzed several simple $S$ matrices. For a given
partial wave $S$ matrix, though very complicated, can be written
as a product of the simple $S$ matrices: \be\label{param}
S^{Phys.}=\prod_iS^{R_i}\cdot S^{cut}\ , \ee since a unitary
matrix divided by any unitary matrix is still unitary. In the
above equation $S^{R_i}$ denotes simple $S$ matrices as described
above and $S^{cut}$ contains only cut which can be parameterized
in the following simple form, { \bqa\label{fS}
 S^{cut}&=&e^{2i\rho f(s)}\nonumber\\
 f(s)&=&f_0+\frac{s-4}{\pi}\int_{L}\frac{{\rm
 Im}_Lf(s')}{(s'-4)(s'-s)}\nonumber\\
 &&+\frac{s-4}{\pi}\int_{R}\frac{{\rm
 Im}_Rf(s')}{(s'-4)(s'-s)}\ ,
\eqa where $R$ denotes cuts at higher energies other than $2\pi$
cuts. The function $f$ must be non-vanishing in general and
especially in $\pi\pi$ scatterings. It should be emphasized that
there is no loss of generality in Eq.~(\ref{param}), since couple
channel effects (or physics at sheet III, IV, etc.) are all hidden
in the right hand cut integral in Eq.~(\ref{fS}). The
Eq.~(\ref{param}) is a re-derivation of the so-called Dalitz--Tuan
parameterization, with the special treatment of $S$ matrix poles
as discussed in sec.~\ref{TfSm}.
\section {The problem of a Breit--Wigner description of the
$\sigma$ resonance}

  A frequently used parameterization form of a resonance
  in the literature is {\be\label{mBW}
S=\frac{M^2-s+i\rho(s)M\Gamma}{M^2-s-i\rho(s)M\Gamma}\ .\ee}
However, such an $S$ matrix contains three poles. For a
sufficiently large $M^2$ and small $M\Gamma$ it contains a
resonance and a \textit{ virtual state}. According to the
discussion made in sec.~\ref{TfSm}, this $S$ matrix can be
factorized in the product of two simpler $S$ matrices. The
Eq.~(\ref{mBW}) contains only two parameters, therefore the pole
location of the virtual state is determined by the pole location
of the resonance. Qualitatively speaking, the virtual state pole
get closer to the threshold when the resonance pole is light and
broad, and vice versa. If denoting the resonance pole position as
$z_0$, then the scattering length of the virtual state is,
{\be\label{scatt_a} a=\frac{{\rm Im}[z_0]} { {\rm
Re}[\sqrt{z_0(z_0-4)}]} \ . \ee} According to the present
parameterization, the scattering length of the virtual state pole
and the resonance pole are additive and are both positive. Taking
$M=400$MeV, $\Gamma=600$MeV as an example ($z_0\equiv
(M+i\Gamma/2)^2$). The scattering length predicted by
Eq.~(\ref{sres}) is 0.23, and is 2.73 as predicted by
Eq.~(\ref{mBW})! It is easy to understand that such a virtual
state pole is not allowed in I=J=0 $\pi\pi$
scattering.~\footnote{In the I=J=2 channel, chiral symmetry
however does predict  a virtual state pole. But the pole locates
very close to $s=0$ and hence has only negligible
effect.\cite{AXZ01}} Because the virtual state pole corresponds to
an $S$ matrix zero on the physical sheet: $S=1+2i\rho T=0$, which
implies $T(s_0)=-1/2i\rho(s_0)$. For $s_0$ not far from the
threshold it predicts $T\sim O(1)$ in chiral power counting, yet
chiral symmetry dictates that $T\sim O(m_\pi^2)$ near threshold.

To use  Eq.~(\ref{mBW}) in studying the $\sigma$ resonance is
therefore in error -- we struggled to look for a pole rather far
from the physical region but end up with a spurious pole much
closer to the physical region and dominates the physics near the
threshold! Unfortunately, this problem is overlooked by most
studies which incorrectly make use of Eq.~(\ref{BW2}).
\section{Conclusions and prospects}
Before jumping to the conclusion I would like to discuss a little
bit more on the background phase in Eq.~(\ref{fS}). One tries to
determine the background term  by a match between our
parameterization and the $\chi$PT results in the region where
$\chi$PT result is reasonable. { \be\label{match}
\prod_iS^{r_i}\cdot S^{cut}\simeq 1+2i\rho (s){\rm T}^{\chi
PT}(s)\ , \ee} where { \be {\rm T}^{\chi PT} =
T^{(2)}+T^{(4)}+T^{(6)}+\cdot\cdot\cdot\ . \ee} On the other side
we have $f=f^{(2)}+f^{(4)}+\cdot\cdot\cdot$ and also $S^{R_i}$ can
be expanded. Taking the simpler I=2,J=0 channel for example (where
there is no resonance pole), up to $O(p^4)$, we have {\bqa
S^{cut}&=&e^{2i\rho f}\simeq 1+2i\rho
(f^{(2)}+f^{(4)})-2\rho^2 f^{(2)2}\nonumber\\
&\simeq& 1+ 2i\rho (T^{(2)}+T^{(4)})\ , \eqa} which leads to
$f^{(2)}=T^{(2)}$,  $f^{(4)}=T^{(4)}-i\rho T^{(2)2}={\rm Re}_R
T^{(4)}$, etc.. This procedure however has a problem by
introducing an essential singularity to the approximate $S$ matrix
at $s=0$.\footnote{Related discussion may be found in
Ref.~\cite{XZ03}.} But the $S$ matrix may be
 acceptable except in the vicinity of $s=0$.
This way of matching with $\chi$PT determines the background phase
without introducing new parameters except those pole parameters,
avoiding the disastrous physical sheet resonance poles predicted
by Pad\'e approximations.\cite{Qin02} Therefore it is worthwhile
to further study along this direction.

To conclude, it is  suggested to use Eq.~(\ref{sres}) to
parameterize the propagator of the $\sigma$ resonance. Notice that
the parameterization form is $not$ in any sense unique. For
example, in production processes, since there are difficulties in
estimating the background contribution, one is free to absorb some
of the background contribution into the propagator. But one thing
is clear, that the spurious virtual state as introduced by
Eq.~(\ref{BW2}) must be removed. There is strong evidence in the
experimental fit to production processes that Eq.~(\ref{sres})
considerably improves the total $\chi^2$ comparing with
Eq.~(\ref{BW2}).\cite{WN}

\acknowledgements

I would like to thank Profs. M.~Ishida, S.~Ishida, K.~Takamatsu,
Y.~S.~Tsai and T.~Tsuru for organizing such a  charming symposium
and for the warm hospitality I received.

\end{document}